\documentclass[11pt,twoside]{article}
\usepackage{asp2006}
\usepackage{psfig}
\usepackage{epsf}
\usepackage{graphics}
\usepackage{graphicx}
\usepackage{lscape}
\begin{document}

\title{The starburst-AGN disconnection}
\author{R. Cid Fernandes$^{1,2}$,
  M. Schlickmann$^{1}$,
  G. Stasi\'nska$^{2}$, 
  N. Vale Asari$^{1,2}$, \\
  J. M. Gomes$^{1,3}$, 
  W. Schoenell$^{1}$,
  A. Mateus$^{4}$, 
  L. Sodr\'e Jr.$^{4}$ \\
  (the SEAGal collaboration)
}
\affil{\footnotesize
  $^{1}$Departamento de F\'{\i}sica, CFM, UFSC, Florian\'opolis, SC, Brazil\\
  $^{2}$LUTH, Observatoire de Paris, CNRS, Universit\'e Paris Diderot, Meudon, France\\
  $^{3}$GEPI, Observatoire de Paris, CNRS, Universit\'e Paris Diderot, Meudon, France\\
  $^{4}$Instituto de Astronomia, Geof\'{\i}sica e Ci\^encias
  Atmosf\'ericas, USP, SP,  Brazil
}

\begin{abstract}
Optical studies of starbursts, AGN and their connections usually leave
out galaxies whose emission lines are too weak to warrant reliable
measurement and classification. Yet, weak line galaxies abound, and
deserve a closer look. We show that these galaxies are either massive,
metal rich star-forming systems, or, more often, LINERs. From our
detailed stellar population analysis, we find that these LINERs have
stopped forming stars long ago. Moreover, their ionizing radiation
field is amazingly consistent with that expected from their old
stellar populations alone.  The black-hole in the centers of these
massive, early-type galaxies is not active enough to overwhelm stellar
ionization, and thus, despite their looks, they should not be called
AGN.
\end{abstract}

\section{Introduction}

\begin{figure}
  \begin{center}
    \includegraphics[bb=30 460 580 700,width=0.86\textwidth]{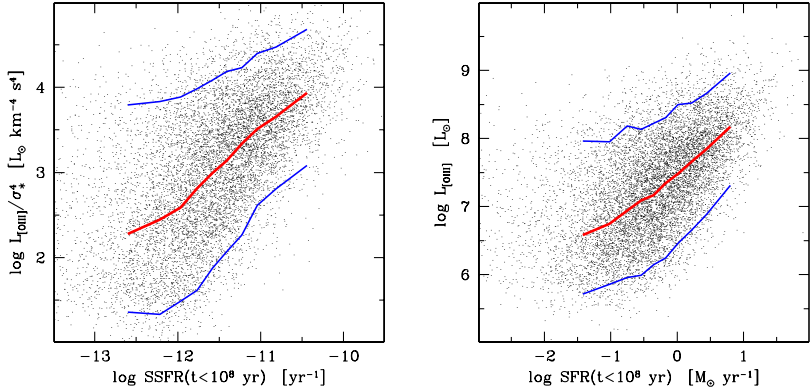}
  \end{center}
  \caption{(a) Relation between the {\sc starlight}-derived specific
SFR and a proxy for the ``specific black-hole accretion rate'' for
$\sim 10000$ type 2 AGN in the SDSS-DR5. All objects lie above the
Kewley et al.\ (2001) maximal-starburst line in the BPT diagram. Lines
mark the 5, 50 and 95 percentiles. (b) SFR versus [OIII] luminosity
(an indicator of accretion rate). Unlike in panel a, the correlation
here is partly induced by distance-dependence.}
\label{fig:SSFR_x_SBMAR}
\end{figure}

The connection between star-formation (SF) and AGN has been the
subject of much work over the past two decades, and this conference
shows that there is still plenty to be done.  There is now solid
evidence that these two phenomena coexist and scale with each other in
at least part of the AGN population. As an example,
Fig.~\ref{fig:SSFR_x_SBMAR} shows the correlation between specific SF
rate and black-hole accretion rate for galaxies classified as type 2
AGN in the SDSS. The current SF rate is computed from our spectral
synthesis analysis with {\sc starlight} (Cid Fernandes et al.\ 2005),
whereas, following current ideas (Heckman et al.\ 2004), $L_{\rm
[OIII]} / \sigma_\star^4$ is used as a proxy for $\dot{M_\bullet} /
M_\bullet$. The strong correlation between these completely
independent tracers of SF and AGN activity is indicative of a
connection, even though a comprehensive and physically sound framework
in which this and other global relations can be understood is still
lacking.

This contribution presents an unconventional perspective on SF-AGN
connections. In fact, we talk more about disconnections! Looking at
plots like Fig.~\ref{fig:SSFR_x_SBMAR}, one may wonder {\em who is
missing?}  Who are the AGN sitting in galaxies which no longer form
stars?  At an even more basic level, how many galaxies are not plotted
there because we do not have enough data to classify them as AGN?

We start drawing attention to the fact that optical studies of
starbursts and AGN normally set aside a huge population of objects
with weak emission lines, inducing potentially dangerous biases.  We
then set out to place this ``forgotten'' population in the context of
current spectroscopic categories.  We show that the majority of these
sources look like low-luminosity AGN (a.k.a.\ LINERs), but can instead
be understood in terms of the retired galaxy model described by
Stasi\'nska et al.\ (2008) and Vale Asari elsewhere in this volume.

\section{The ``forgotten'' population of weak line galaxies in the SDSS}

\begin{figure}
  \begin{center}
    \includegraphics[bb=30 200 580 700,width=0.54\textwidth]{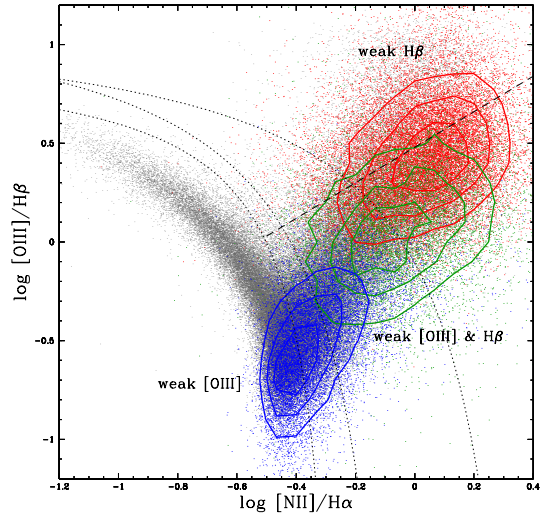}
  \end{center}
  \caption{BPT diagram for galaxies with $S/N > 3$ in at least
H$\alpha$ and [NII]. Grey dots correspond to objects with $S/N > 3$ in
all BPT lines. Blue, green and red dots and contours correspond to
WLGs of different types. Dotted curves mark the SF/AGN borderlines of
(from left to right) Stasi\'nska et al.\ (2006), Kauffmann et al.\
(2003) and Kewley et a.\ (2001), while the diagonal line is the
Seyfert/LINER borderline derived by Schlickmann (2008) by means of
optimal separation techniques applied to the (more complex and
``expensive'') Kewley et al.\ (2006) classification scheme.}

\label{fig:BPT_WLGs}
\end{figure}

In an era where surveys flourish, and this very volume is a testimony
of this trend, galaxies are often tagged as ``starburst'' or
``active'' on the basis of relatively little data. Even worse, in many
cases the data is not enough to warrant a convincing classification,
and such objects are usually set aside.

In the case of the SDSS, one has to make do with optical spectra,
classifying galaxies on the basis of their location in diagnostic
diagrams such as the classical [OIII]/H$\beta \times$ [NII]/H$\alpha$
BPT diagram.  The SDSS is an exquisite data set, but an awful lot of
its galaxies do not have enough $S/N$ in their emission lines to allow
a reliable classification. To illustrate the size of this problem, we
select galaxies where both H$\alpha$ and [NII]$\lambda6584$ are
detected with $S/N > 3$. Of this population of emission line galaxies,
about 1/5 do not have [OIII]$\lambda5007$ and/or H$\beta$ detected
with the same level of confidence, such that a rigorous BPT
classification is not possible. This may not seem so much, but it so
happens that this fraction increases to over $1/2$ counting only
objects where $\log {\rm [NII]/H\alpha} > -0.2$, a criterion which
completely excludes pure SF systems.  These objects thus pertain to
the ``AGN-wing'' in the BPT diagram, but are they LINERs, Seyferts, or
SF$+$AGN composite systems?  Clearly, one would like to have a better
idea of what these galaxies are, as neglecting such a numerous
population may introduce severe biases, with potentially hazardous
consequences.

An unorthodox but statistically valid way to investigate the nature of
these weak line galaxies (WLGs) is to plot them in the BPT diagram
irrespective of the $S/N$ of their emission lines. This is done is
Fig.~\ref{fig:BPT_WLGs}, where the grey points correspond to galaxies
with $S/N > 3$ in H$\beta$, [OIII], H$\alpha$ {\em and}
[NII]. Contours are used to indicate the location of WLGs, split in
those with: (i) $S/N > 3$ in H$\beta$ but not in [OIII], (ii) $S/N >
3$ in [OIII] but not in H$\beta$, and (iii) those where neither
H$\beta$ nor [OIII] has $S/N > 3$.  Galaxies with weak [OIII] but
strong H$\beta$ are located at the bottom of the SF wing, where metal
rich SF systems are expected to be, intruding a little bit into the
AGN wing. These are the most massive and metal rich SF galaxies. WLGs
of type (ii) and (iii), however, are some other sort of thing. They
are well within the AGN wing, heavily skewed towards its low
excitation, LINER-like branch.  The diagonal line in
Fig.~\ref{fig:BPT_WLGs} maps the Seyfert/LINER separation of Kewley et
al.\ (2006), which requires [OI] and [SII] lines, to the BPT plane.
In her MSc thesis, Marielli Schlickmann proposes alternative methods
to rescue WLGs from the classification limbo. Her more robust results
confirm the overall picture sketched above in Fig.~\ref{fig:BPT_WLGs}:
Discounting weak-[OIII] sources, most WLGs look like LINERs.

LINERs are thus much more common than one would think selecting only
galaxies with decent $S/N$ in all BPT lines.  But {\em what are
LINERs?}  Despite early warnings that they constitute a mixed bag
(Heckman 1987), ``LINER'' and ``low luminosity AGN'' are basically
synonyms in the current literature. Sure enough, many nearby, well
studied LINERs are truly {\em Active} Galactic {\em Nuclei}, in the
sense that they have accreting nuclear black-holes. LINERs in the
SDSS, including the weak-line ones described above, are all massive
and early-typish, and thus should also harbor super-massive
black-holes. But are these black-holes truly {\em active}? In other
words, are they responsible for the emission lines we see?

\section{The starburst-AGN disconnection: LINERs as retired galaxies}

\begin{figure}
  \begin{center}
    \includegraphics[bb=45 490 580 710,width=0.99\textwidth]{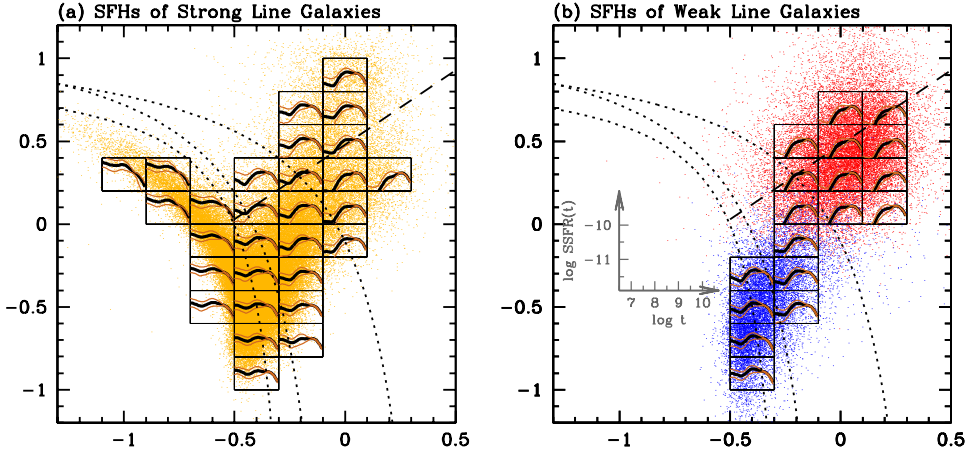}
  \end{center}
  \caption{{\sc starlight}-derived star-formation histories of (a)
strong and (b) weak line galaxies plotted on the BPT diagram. Each box
shows the 16, 50 and 84\% percentiles of the $t$-by-$t$ specific SFR
(see inset for scale). In (b), blue corresponds to weak [OIII] but
strong H$\beta$, and vice-versa in red.}
\label{fig:SFHs_on_BPT}
\end{figure}

\begin{figure}
  \begin{center}
    \includegraphics[bb=30 445 580 700,width=0.81\textwidth]{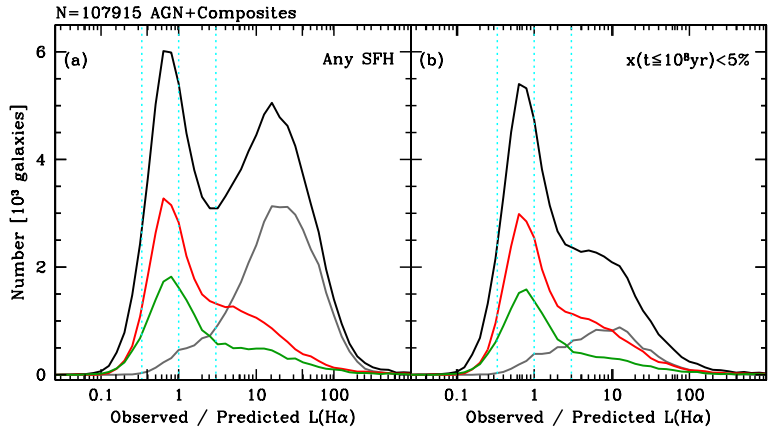}
  \end{center}
  \caption{Ratio between the observed H$\alpha$ luminosity and that
predicted counting only ionization by post-AGB stars for a sample of
over $10^5$ non-SF galaxies. Grey lines are for strong line sources
($S/N > 3$ in all BPT lines). Red and green lines are WLGs (as in
Fig.\ \ref{fig:BPT_WLGs}). Panel (a) includes all galaxies, while in
(b) only those without significant recent SF are plotted.}
\label{fig:Xi_histograms}
\end{figure}

Further insight can be gained by inspecting the star-formation history
(SFH) of galaxies across the BPT diagram. We have done this several
times before, but now we include WLGs.  Fig.~\ref{fig:SFHs_on_BPT}a
shows the specific SF rate as a function of lookback time for galaxies
with $S/N > 3$ in all four BPT lines. The balance of current to past
SF shifts gradually from the top of the SF wing to its bottom, and
there are also differences along the AGN wing, both vertically and
horizontally, with Seyferts having significantly more recent SF than
LINERs. Fig.~\ref{fig:SFHs_on_BPT}b shows the SFHs of WLGs, again
using the dangerous but statistically valid trick of computing
[OIII]/H$\beta$ using data where at least one of these lines has very
low $S/N$. Galaxies with weak [OIII] but ``strong'' ($S/N > 3$)
H$\beta$ all show significant ongoing SF, confirming that they are
indeed SF systems. Those with weak H$\beta$ but strong [OIII],
however, have essentially no SF in the past $\sim 10^8$ yr. Their SFHs
are similar to those of LINERs in Fig.~\ref{fig:SFHs_on_BPT}a, but
with even less SF in the recent past.  Star-birth, therefore,
cannot be responsible for any significant part of the ionizing
radiation field in LINERs. But what about ``{\em star-retirement}''?

Trincheri \& di Serego Alighieri (1991) and later Binette et al.\
(1994) proposed that post-AGB stars and white dwarfs could be
responsible for the gas ionization in elliptical galaxies.
Stasi\'nska et al.\ (2008) revisited this idea, and, with the aid of
the stellar populations derived by {\sc starlight}, produced
self-consistent models which show that indeed many LINERs can be
explained in this way. In that paper we required good $S/N$ in all BPT
lines to ensure reliable data and classification. With this standard
quality control, and using Bruzual \& Charlot (1993) models, $\sim
1/4$ of SDSS LINERs are compatible with being retired galaxies whose
gas is ionized by old stars.  Fig.\ \ref{fig:Xi_histograms} shows that
this fraction increases a lot including WLGs.  The x-axis in these
histograms is the ratio of the observed H$\alpha$ luminosity to that
expected from post-AGB and white dwarfs alone. Only galaxies above the
Kauffmann SF/AGN line are included in this exercise. The plot shows
that most WLGs are consistent with the retired galaxy model to within
a factor of 3. Most strong line galaxies, on the other hand, require
far more ionizing photons than old stars can provide. This bona-fide
AGN population shrinks dramatically in number when filtering out
galaxies where over 5\% of the light comes from populations younger
than $10^8$ yr (Fig.\ \ref{fig:Xi_histograms}b), where the
``starburst-AGN connection'' is in full swing.  The bimodality in
Fig.\ \ref{fig:Xi_histograms} is much less pronounced among systems
dominated by old stellar populations.

To summarize, many (actually: most, if WLGs are not ignored)
LINER-looking systems in the SDSS are consistent with being retired
galaxies, whose ionizing photons are produced by ageing stars.  This
is not to say that they do not contain a massive black hole, but until
the black hole is proven to be active by means of independent data,
they should not be included in any census of actively accreting
black-holes.

\end{document}